**Characterizing climate predictability and model response variability from multiple initial condition and multi-model ensembles**

Devashish Kumar and Auroop R. Ganguly[*]

Sustainability and Data Sciences Laboratory (SDS Lab),
Department of Civil and Environmental Engineering,
Northeastern University, Boston, MA-02115, USA
Email: a.ganguly@neu.edu
Phone: 617-373-6005

**Abstract**

Climate models are thought to solve boundary value problems unlike numerical weather prediction, which is an initial value problem. However, climate internal variability (CIV) is thought to be relatively important at near-term (0-30 year) prediction horizons, especially at higher resolutions. The recent availability of significant numbers of multi-model (MME) and multi-initial condition (MICE) ensembles allows for the first time a direct sensitivity analysis of CIV versus model response variability (MRV). Understanding the relative agreement and variability of MME and MICE ensembles for multiple regions, resolutions, and projection horizons is critical for focusing model improvements, diagnostics, and prognosis, as well as impacts, adaptation, and vulnerability studies. Here we find that CIV (MICE agreement) is lower (higher) than MRV (MME agreement) across all spatial resolutions and projection time horizons for both temperature and precipitation. However, CIV dominates MRV over higher latitudes generally and in specific regions. Furthermore, CIV is considerably larger than MRV for precipitation compared to temperature across all horizontal and projection scales and seasons. Precipitation exhibits larger uncertainties, sharper decay of MICE agreement compared to MME, and relatively greater dominance of CIV over MRV at higher latitudes. The findings are crucial for climate predictability and adaptation strategies at stakeholder-relevant scales.

**Introduction**

Climate adaptation stakeholders such as water resource managers[1,2], ecosystem conservationists[3], and infrastructures owners and operators[4] are increasingly seeking reliable information on regional climate change at near-term planning horizons[5–7]. Such information is available from climate models; however, inadequate understanding and improper communication[8] of the associated uncertainty are often a limiting factor in utilizing that information in decision-making[1–3,6,9]. Recently Mora *et al.*[10] presented timing of emergence of signal of regional climate change, which is important for adaptation planning[11]; nevertheless, lack of consideration of uncertainty especially internal variability (natural fluctuations of the climate system in the absence of external forcing) led to imprecise estimates[11].

Weather forecasting, where both predictive skills and predictability[12] inform decisions, provides an analogy. While numerical weather prediction models have improved[13] significantly over the years, chaos theory suggests limits[14] to predictability. The tradeoff between predictive skills versus predictability in weather forecasts implicitly guides impacted systems such as air traffic, logistics movements, response to impending weather hazards, and transfer of risk using financial instruments such as weather derivatives[15]. No such guidance exists in climate[3], which severely limits risk management and mitigation in the context of adaptation.

By convention[14], climatology is often defined as 30-year average of the weather. Thus, climate model simulations are often averaged over 30-year windows (although decadal averages are also occasionally used). For a specified greenhouse gas emissions-scenario, uncertainty in projected climate change primarily results due to climate internal variability (CIV) and model response variability (reflecting our lack of knowledge of the climate system or the inability to encapsulate the existing knowledge within climate models). Model response variability (MRV) is caused by structural or parametric differences across multiple climate models. CIV typically dominates climate uncertainty in the first few decades of projection horizons[12]. Although longer-term climate modeling is thought to be a boundary-condition rather than an initial-condition problem, CIV continues to be one of the dominating factors in overall projection uncertainties over 0-30 year ("near-term") planning horizons. Adaptation studies typically use a 30-year average. However, when stakeholders' require planning horizons to be near-term, the influence of CIV cannot be assumed to be negligible a priori[3].

Two approaches (see Methods) have been suggested to characterize CIV from the Coupled Model Intercomparison Project (CMIP) model projections in the prior literature. The first[16,17] does not use MICE but relies on just one run from any specific model. A trend line (linear[18] or a fourth-order polynomial[16,17]) is fitted to that single model run and the variance of the deviations from that trend line is used as an estimate of CIV specific to that model (following which CIV estimates are averaged across all models to provide an overall estimate). The second approach[8,19] does use MICE and relies on the deviations of individual initial condition (IC) runs from the average of all MICE runs. The focus of the latter[8,19] was to examine the deviations in projections resulting from different initializations.

Decadal climate predictions[7,12], a relatively new line of research, which is still in infancy, may eventually be able to address the challenges in characterizing CIV. However, this line of research is barely emerging and sufficient number of decadal prediction runs from multiple models is not yet available. Once adequate model ensembles for decadal predictions become available, their skill will need to be evaluated systematically through hypothesis-driven analysis prior to their use for adaptation.

The presence of a (relatively) large set of MICE[20] enables us for the first time to formulate and examine a research question: How does CIV (or MICE agreement) compare to MRV (or MME agreement) based on a direct comparison over decadal (0-30 year) to century scales across multiple spatial resolutions? While the problem is not dissimilar to the one examined by Hawkins and Sutton[16,17], our analysis can be designed as a sensitivity analysis experiment. Specifically, we can directly compare MICE runs of the single model (similar to generating multiple IC runs by holding one model constant) to MME runs based on a single initialization. We examine the hypothesis that a direct comparison of CIV versus MRV provides the same insights as those obtained from Hawkins and Sutton[16,17] and Räisänen[18]. Furthermore, we examine the hypothesis that different variables (specifically temperature and precipitation) may exhibit considerable differences in the relative dominance of CIV versus MRV over 0-100 years across multiple (local to global) spatial resolutions. Previous assessments[12,17,18] reported that CIV for temperature dominates over projection lead times of one to three decades and MRV thereafter at all spatial scales for a particular emissions-scenario. Similar assessments[12,16,18] for precipitation concluded that MRV is the dominant source at all time horizons although CIV is important in the first few decades.

The MICE simulations used in this study are the same as Kay *et al*[20]. Specifically we use the Community Earth System Model (CESM)[20] model, which recently ran simulations using one climate model (CESM-CAM5) but for 30 different ICs for historical (1920-2005) and Representative Concentration Pathways 8.5 (RCP8.5) forcing (2006-2100). The MME simulations were obtained from the Coupled Model Intercomparison Project phase 5 (CMIP5) models[21] (Table S1). As discussed, the availability of the recent MICE data provides us an opportunity to perform sensitivity analysis to understand the relative importance of different sources of uncertainty by directly quantifying CIV (30 ICs, one model)[20] and MRV (30 models, one IC)[21].

**Results**

Figure 1 shows that MICE versus MME runs may not be visually distinguishable either for temperature or for precipitation. Spatial patterns of change in surface warming and precipitation for MICE (Figs 1c and 1k) and MME (Figs 1g and 1o) medians look similar. Most of the individual realizations from both MICE (Figs 1a and 1b) and MME (Figs 1e and 1f) runs show varying degree of warming patterns over most regions of North America (Figs S1.1-S1.4). Uncertainty, quantified as interquartile range (IQR)[22], in projecting regional warming is less from MICE (Fig. 1d) runs than from MME (Fig. 1h). Compared with temperature, we observe strikingly different spatial patterns over western United States and parts of Canada for projection of regional precipitation from MICE (Figs 1i and 1j) and MME (Figs 1m and 1n) simulations. The individual realizations from multi-initial conditions and multi-models show larger spatial variability for precipitation (Figs S1.5-S1.8)) than for temperature (Figs S1.1-S1.4). The CIV (Fig. 1l) and MRV (Fig. 1p) may be visually compared. The central tendency of a set of ensembles (whether MICE [Figs 1c and 1k] or MME [Figs 1g and 1o]) may not convey adequate information to stakeholders since the plausible futures with hotter/cooler or drier/wetter regions may be underemphasized.

Figure 2 shows relative agreement among MICE and MME runs in projecting annual and seasonal climate change as a function of horizontal scales. Relative agreement, calculated as in Räisänen (2001)[18], is a dimensionless number that varies between 0 and 1. A value of 1 indicates perfect agreement among ensemble members in projecting climate change and that of 0 indicates climate change is totally random. Relative agreement is high for temperatures even at local scales and increases for precipitation with spatial scales for all projection time horizons. The agreement among MICE runs is more than MME runs for both temperature and precipitation change at all horizontal scales for all lead times. The agreement is highest at the global scale and lowest at grid-box scale for both MICE and MME runs; however, it declines sharply for multi-model runs. For precipitation, multi-model agreement is low at grid-box and regional scales for near-term (Fig. 2, top right); nevertheless it increases slightly at the end of this century (Fig. 2, bottom right). On the other hand, the agreement among MICE runs increases sharply for projection time horizons of mid-to-end of the century.

Figure 3 illustrates spatial patterns of relative agreement at grid-box scales for annual and seasonal means of change in temperature and precipitation over near-term horizons (2010-2039). For temperature, the local agreement is high almost everywhere (Fig. 3) and is perfect towards the end of the century (Figs S3.1-S3.2) for MICE runs (Figs 3, S3.1-S3.2; 1$^{st}$ column) and is relatively high over topics compared to extratropics for MME runs (Figs 3, S3.1-S3.2; 2$^{nd}$ column). For precipitation, local agreement is low compared to temperature for both MICE and MME ensembles. For precipitation, the relative agreement is low for MME runs (Figs 3, S3.1-S3.2; 4$^{th}$ column) over tropics compared to extratropics; it reflects our inadequate understanding of physical processes over tropics.

Figure 4 shows projections of global mean surface temperature and precipitation for decadal means of annual and seasonal means. The anomaly time series, relative to 1961-1990, is shown for each of the individual IC and model simulations along with their ensemble medians. For temperature, the width of CIV remains constant, while it keeps widening for MRV with projection lead times (2006-2099) at all resolutions. The contribution of CIV to total uncertainty decreases as climate change signal emerges with lead times. For temperature, MRV is comparable to CIV for projection time horizons of one to two decades; however, MRV dominates beyond 2030. The dominance of MRV is also observed over global land, tropics, extratropics, and grid-box scales (Figs S4.1-S4.10). The CIV for temperature is always contained within the multi-model spread at all resolutions (Figs S4.1-S4.10; left). For global mean precipitation, CIV is comparable to MRV until about mid-century. MRV dominates beyond mid-century however the dominance is not as strong as it is for temperature. For precipitation, the influence of CIV is stronger for all lead-times at higher spatial resolutions (Figs S4.1-S4.10; right).

Figure 5 shows the fraction of total variance in mean temperature and precipitation projections due to MRV (blue) and CIV (orange) with projected lead times (2006-2099) at multiple spatial scales. For annual and seasonal mean change in temperature, the relative contribution of CIV decays sharply, and MRV is the dominant sources of uncertainty for all projection time horizons at all spatial scales (Figs S5.1-S5.2; top two rows). However, CIV is comparable to than MRV for projection of precipitation for global, regional, and grid-box scales (Figs S5.1-S5.2; bottom two rows).

Figure 6 displays spatial maps of fraction of total variance explained by CIV during boreal summer for four decades. MRV contributes more to climate uncertainty for projected changes in annual and seasonal mean temperature over land (Figs 6, S6.1-S6.2; right) and ocean (Figs S6.3-S6.5; right) for all projection time horizons. However for precipitation over land, CIV dominates climate uncertainty especially at higher latitudes, where climate change hotspots are expected to emerge sooner[23]. In addition, CIV dominates over Australia, Southern Africa, Amazonia, and Southern South America during JJA and over India, Sahel during DJF for all projection lead times (Figs 6, S6.1-S6.2; left). For precipitation over ocean (Figs S6.3-64.5; left), there is no uniform pattern of dominance of either sources of uncertainty; nonetheless the spatial extent of contribution of MRV increases with projection lead times.

**Discussion**

Two key findings emerge from our study. First, for changes in precipitation, CIV is comparable to MRV over decadal to century scale projection horizons. Second, for both temperature and precipitation, relative agreement for MICE is considerably larger than MME across local to global horizontal scales for all lead times. The implication of the first finding relates to the treatment of different sources of uncertainty in planning or design principles[3]. Specifically, likelihood based risk formulations[24] may not necessarily be appropriate for situations in which CIV dominates, and flexibility in design and planning may be necessary. The second finding suggests the need for continued focus on model development[25] and methods for diagnosis[26,27] and prognosis[28] of multi-model simulations. Regions and seasons where MRV dominates climate uncertainty, there is scope to reduce the uncertainty further by investment in model development[26,29,30]. High multi-model agreement or reduced MRV over high latitudes corroborates our improved understanding of atmospheric process in these regions.

The MRV-to-CIV Ratio (MCR) and Relative-Agreement (MCRA) behave as expected[12] for temperature across spatial scales and projection horizons, specifically, MCR (MCRA) is low (high) other than specific regions over high latitudes and/or in the near-term. However, the dominance of CIV for northern Indian summer temperatures is unexpected. For precipitation, MCR is relatively high under situations where the physics is less well understood, specifically the tropics, summer precipitation, South Asia, Sahel, Australia, South Africa and parts of South America. These reflect, respectively, lack of understanding over tropics[29] (especially the double inter-tropical convergence zone[31] and the equatorial Pacific cold tongue[31]) and of organized deep convection[29], the South Asian monsoon[32], Sahel[33], and low-level tropical jets in Amazonia[34]. However, an interesting finding is the higher absolute value of precipitation CIV over the tropics relative to the extratropics. The CIV dominance for precipitation over higher latitudes, while important for policy, is expected. However, regional dominance, especially at decadal to mid-century horizons, of precipitation CIV in Australia, central South America, Brazilian coastlines, Sahel, India, Southern Africa and parts of North America in the winter, as well as over northern Africa and the middle East in the summer, are unexpected but interesting.

**Methods**

**Characterization of Climate Internal Variability:** Prior literature suggests two possible approaches for characterizing climate internal variability (CIV). The first approach was proposed by Hawkins and Sutton[16,17] and has been adopted[12] by the Intergovernmental Panel on Climate Change (IPCC). This approach does not rely on multiple initial condition (IC) runs. First, smoothed trend line (specifically, a fourth-order polynomial) is fitted to one single run of any given model. Second, a measure of dispersion (specifically, the variance) is calculated for the fluctuations (or, deviations) from the trend line. Thus, CIV estimates are obtained on a per-model basis from just one IC run for each of the models. Furthermore, the CIV estimates from each model are averaged (weighted by the historical skills of each model) to obtain an overall CIV for the climate system as indicated by the ensemble of models. A variant of this approach (which used best-fit linear trend line instead of a fourth-order polynomial) for estimating CIV was also used[18] in the context of scale dependence of the relative agreement among model ensembles. This basic approach was originally developed for the Coupled Model Intercomparison Project Phases 2 and 3 (CMIP2/CMIP3) model ensembles, which in turn were used by the IPCC for their third and fourth[35] assessment report (TAR/AR4) respectively. We note the following statements in one of prior publication[18]: "*Estimating the magnitude of internal variability would require, in principle, that each CMIP2 model had been used to make several similar $CO_2$ experiments with different initial conditions*". The second approach proposed by Deser *et al.*[8] did use multiple IC ensembles. Internal variability in this case was defined based on deviations of individual IC runs from the average of all IC runs for a specific model. The focus of this work was to examine the deviations of individual IC runs at given space-time locations. Two versions of the NCAR (National Center for Atmospheric Research) model were used to generate multiple IC runs for this purpose. The two versions were the older Community Climate System Model version 3 (CCSM3)[8,19] and the more recently developed Community Earth System Model-Community

Atmospheric Model version 5 (CESM-CAM5)[20]. Our approach relies on multiple IC runs (from a single model: CESM-CAM5) as in Deser et al.[8] but our focus is to obtain estimates of CIV as in Hawkins and Sutton[16,17]. Our research is motivated by the need to compare and contrast model response variability (MRV) versus CIV. The recent availability of the 30 IC ensembles[20] enables design of experiment where we can perform sensitivity analyses by holding a model (CESM-CAM5) constant and examining CIV followed by holding an initialization constant and looking at MRV.

**Relative Agreement:** An approach for characterizing the relative agreement across multi-model and multi-initial condition ensembles, as a function of horizontal scales, was proposed by Räisänen[18], originally examined on CMIP2 ensembles and eventually adopted by the IPCC AR4[35]. The approach computed a relative agreement metric, which considered both the "climate change signal" and the "with-in sample variability" (defined next). Climate change is computed as the difference between 30-year average of projected (RCP8.5 scenario) and simulated (historical) climatology. For temperature, climate change was calculated as the absolute difference (in degree Centigrade), and for precipitation, it was expressed as percentage change with respect to the past climatology (1961-1990). The average squared amplitude of climate change was partitioned into a common signal and variances associated with internal or multi-model variability. The relative agreement ($F$) is mathematically expressed as $F = (M^2/A^2)$, where $M^2$ and $A^2$ were defined as $M^2 = m^2 - (1/k - 1) e^2$ and $A^2 = m^2 + e^2$. Here, $m^2, e^2$, and $k$ denote squares of mean climate change and sample variance, and the sample size (here number of IC or model runs, which is 30) respectively. The value of $F$ varies between 0 and 1. The value $F = 1$ indicates perfect agreement among ensemble members in projecting climate change, and $F = 0$ indicates projected climate change is random. In Figs 3 and S3.1-S1.2, $F$ is computed at grid-box scale to show the spatial variation of local relative agreement. In Fig. 2, $F$ is computed at multiple horizontal scales. To compute the value of F as a function of horizontal scales, the local value of projected climate change was replaced by the weighted area mean in neighboring circular region with diameter, D where the following values of D were considered: 1250, 2500, 5000, 10000, 20000 [hemispherical], and 40000 km [global]). Subsequently, F is computed as the ration of area means of $M^2$ and $A^2$ (see Räisänen[18] for details).

**Climate data:** We obtained monthly temperature and precipitation data for 30 initial condition runs from the National Climate Atmospheric Research (NCAR) community earth system model (CESM-CAM5)[20]. We also retrieved monthly historical and projected (RCP8.5) surface air temperature (*tas*) and precipitation (*pr*) data from 30 CMIP5 models[21] listed in Table S1. The exact number of CMIP5 models (30) was selected as the number of initial condition runs to perform the sensitivity analysis. The selection of a particular climate model from the CMIP5 archive was based on its horizontal grid size; models with finest grid size were selected. For each model, we used only one initial condition realization (r1i1p1). Data were bi-linearly interpolated from models' native grid to 2-degree.

**Acknowledgements**

The research was funded by the U.S. National Science Foundation's (NSF) Expeditions in Computing award #1029711, NSF CyberSEES award #1442728, and NSF BIGDATA award # 1447587. We thank Jouni Räisänen for sharing his code to compute relative agreement metric. We used Climate Data Operators (CDO), the R Language for Statistical Computing, and National Aeronautics and Space Administration (NASA) Panoply for data processing, analysis, and visualization. CMIP5 model data were obtained from the Program for Climate Model Diagnosis and Intercomparison (PCMDI) website. Multiple initial condition data were obtained from the National Center for Atmospheric Research (NCAR) Community Earth System Model (CESM) Large Ensemble Community project.

**Author Contributions**

Both authors did the research and wrote the paper.

**Competing Financial Interests**

Authors declare no competing financial interests.

**Figure Legends**

**Fig. 1. Spatial patterns of surface warming and precipitation change.** Top two rows illustrate projected changes in mean temperature (°C) over North America during 2010-2039, relative to 1961-1990, for June-July-August (JJA) for selected initial condition realizations (#8 [**a**], #13 [**b**], and multi-initial condition median [**c**]) and for selected CMIP5 models (CanESM2 [**e**], GISS-E2-R [**f**], and multi-model median [**g**]). The variability (°C), measured as interquartile range (IQR), across multi-initial condition (MICE) and multi-model (MME) ensembles is shown in **d** and **h**, respectively. Bottom two rows show projected changes in summer mean precipitation (%) for the same set of initial conditions (#8 [**i**], #13 [**j**], and MICE median [**k**]) and of models (CanESM2 [**m**], GISS-E2-R [**n**], and MME median [**o**]) as the first two rows, and the corresponding variability (%) is shown in **l** and **p**. Patterns of change in precipitation for 30 initial conditions and 30 models, including ensemble mean and median, are shown in Figures S1.1-S1.8 for JJA and December-January-February (DJF).

**Fig. 2. Relative agreement at multiple spatiotemporal scales.** Relative agreement among MICE (solid) and MME (dotted) runs for projected changes in temperature (left) and precipitation (right) for annual (ANN), DJF, and JJA means as a function of horizontal scale ("Loc": grid-box scale; "Hem": hemispherical scale; "Glob": global mean). Statistics have been computed, relative to 1961-1990, for three projection horizons: near-term (2010-2039; top), intermediate-term (2040-2069; middle), and long-term (2070-2099; bottom).

**Fig. 3. Spatial maps of relative agreement.** Spatial patterns of relative agreement among MICE and MME runs for projected changes in temperature (left two panels) and precipitation (right two panels), during 2010-2039 relative to 1961-1990, for ANN, DJF, and JJA means.

**Fig. 4. Uncertainty in projection of change in temperature and precipitation.** Projections of global mean surface air temperature (left) and precipitation (right), relative to 1961-1990, for decadal annual (top) and seasonal means (DJF [middle] and JJA [bottom]) from 30 initial conditions (orange) and 30 CMIP5 models (blue). Each individual thin line represents one realization of future climate, and the thick lines represent MICE (orange) and MME (blue) means. Projections from the common model (CESM-CAM5) are shown in green.

**Fig. 5. Relative dominance of different sources of uncertainty.** The fraction of variance explained by model response variability (blue) and climate internal variability (orange) for decadal JJA mean temperature (top two rows) and precipitation (bottom two rows) at multiple

spatial aggregations (Global, Global Land, NHEX, Tropics, North America, and United States) and a few selected cities (Phoenix, Seattle, New York, and Mazatlan).

**Fig. 6. Spatial maps of fraction of total variance.** Spatial maps of fraction of total variance (%) explained by climate internal variability for boreal summer (JJA) precipitation (left) and temperature (right) for the first (2010-2019), third (2030-2039), fifth (2050-2059), and ninth (2090-2099) decade over land. Shades of red highlight dominance of internal variability over model response.

**Supplementary Information**

Table S1. List of CMIP5 models with their horizontal resolution

Fig. S1.1. Projected changes in mean temperature (°C) during 2010-2039, relative to 1961-1990, for DJF for each of 30 initial conditions along with ensemble mean and median
Fig. S1.2. Same as Fig. S1.1 but for JJA
Fig. S1.3. Projected changes in mean temperature (°C) during 2010-2039, relative to 1961-1990, for DJF for each of 30 CMIP5 models along with multi-model mean and median
Fig. S1.4. Same as Fig. S1.3 but for JJA
Fig. S1.5. Projected changes in mean precipitation (%) during 2010-2039, relative to 1961-1990, for DJF for each of 30 initial conditions models along with ensemble mean and median
Fig. S1.6. Same as Fig. S1.5 but for JJA
Fig. S1.7. Projected changes in mean precipitation (%) during 2010-2039, relative to 1961-1990, for DJF for each of 30 CMIP5 models along with multi-model mean and median
Fig. S1.8. Same as Fig. S1.7 but for JJA
Fig. S3.1. Same as Fig. 3 but for 2040-2069
Fig. S3.2. Same as Fig. 3 but for 2070-2099

Fig. S4.1. Same as Fig. 4 but for Global Land
Fig. S4.2. Same as Fig. 4 but for Northern Hemisphere Extratropics (NHEX)
Fig. S4.3. Same as Fig. 4 but for tropics (TROP)
Fig. S4.4. Same as Fig. 4 but for Southern Hemisphere Extratropics (SHEX)
Fig. S4.5. Same as Fig. 4 but for North America
Fig. S4.6. Same as Fig. 4 but for United States
Fig. S4.7. Same as Fig. 4 but for Phoenix
Fig. S4.8. Same as Fig. 4 but for Seattle
Fig. S4.9. Same as Fig. 4 but for New York
Fig. S4.10. Same as Fig. 4 but for Mazatlan
Fig. S5.1. Same as Fig. 5 but for annual decadal mean
Fig. S5.2. Same as Fig. 5 but for DJF decadal mean

Fig. S6.1. Same as Fig. 6 but for annual mean over land
Fig. S6.2. Same as Fig. 6 but for DJF over land
Fig. S6.3. Same as Fig. 6 but for annual mean over ocean
Fig. S6.4. Same as Fig. 6 but for DJF over ocean
Fig. S6.5. Same as Fig. 6 but for JJA over ocean

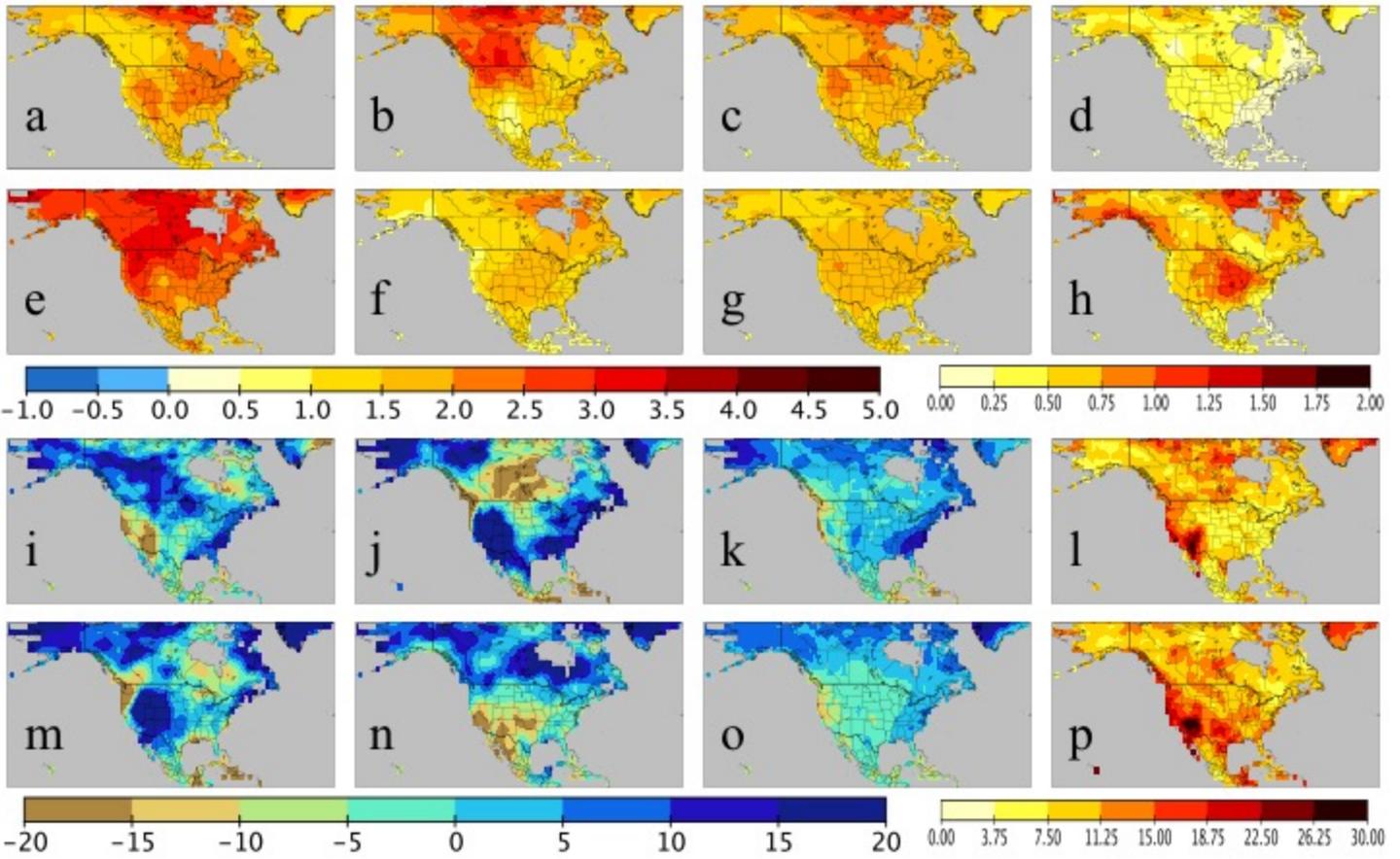

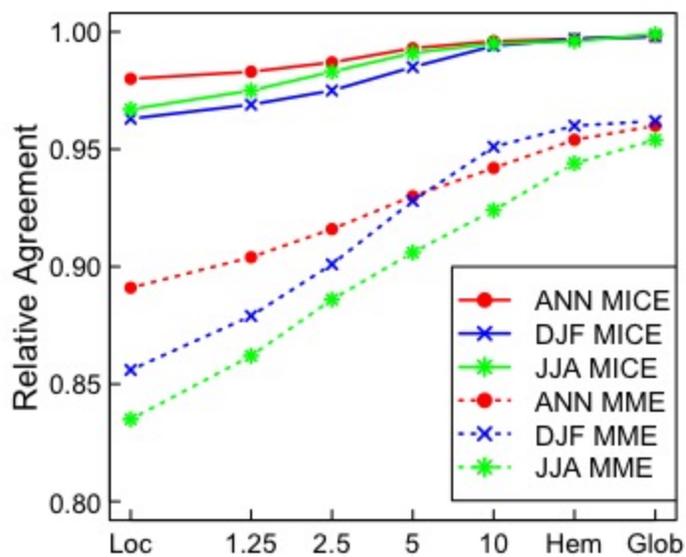
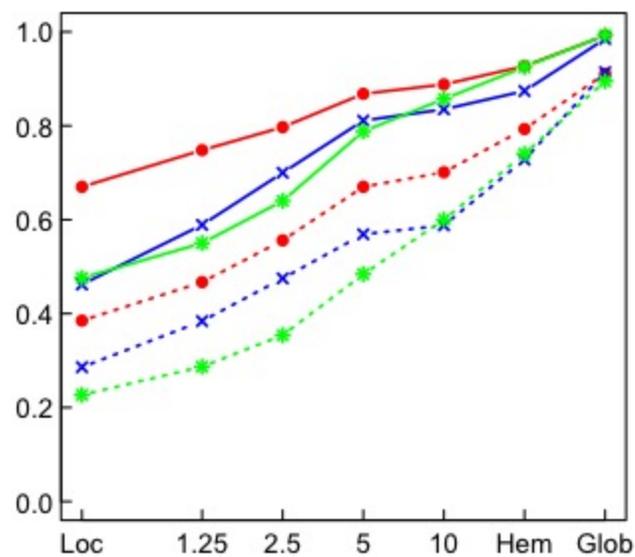
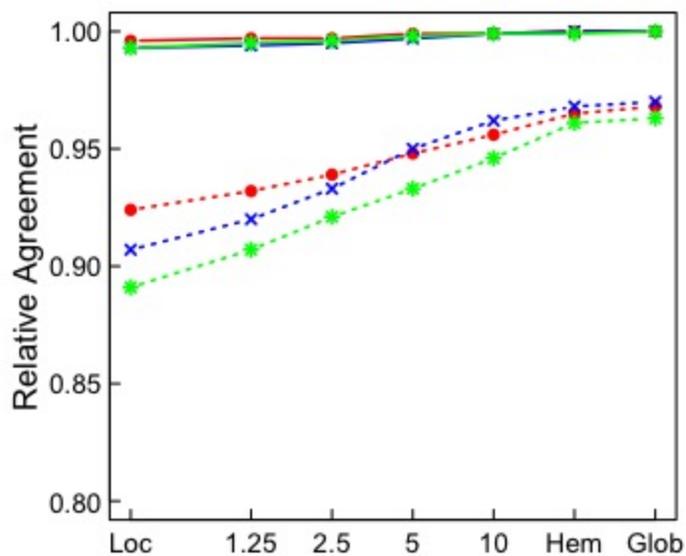
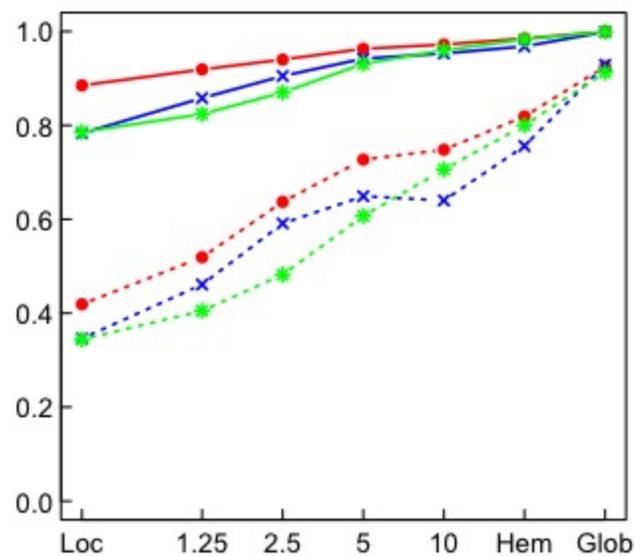
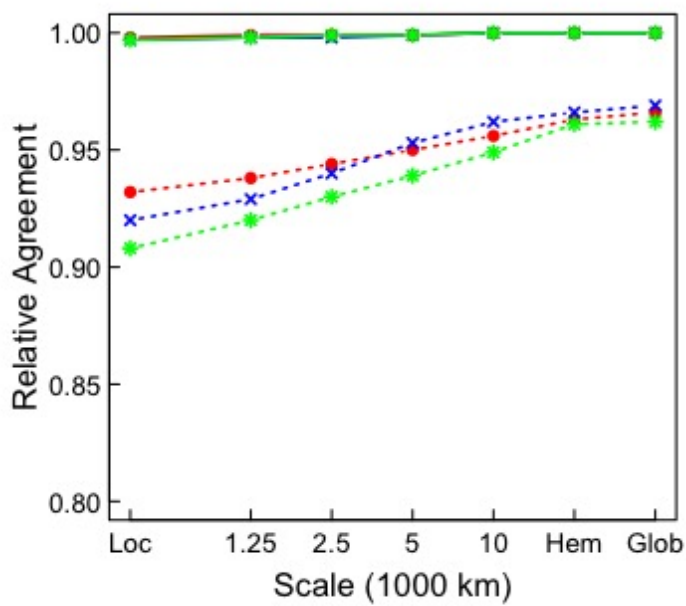
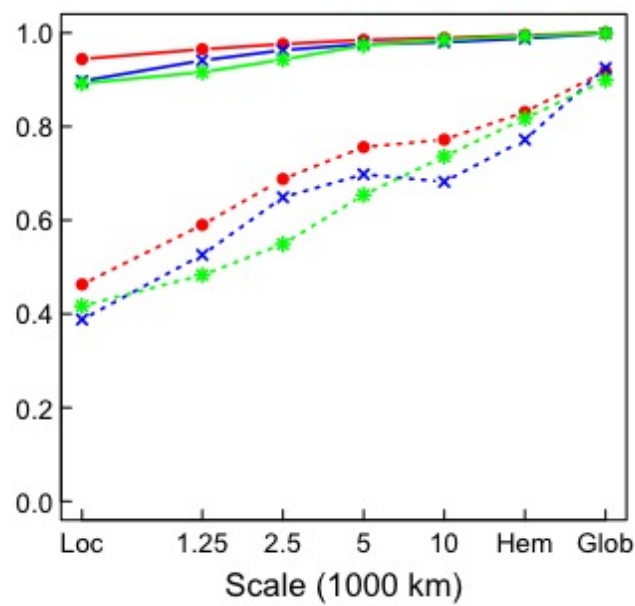

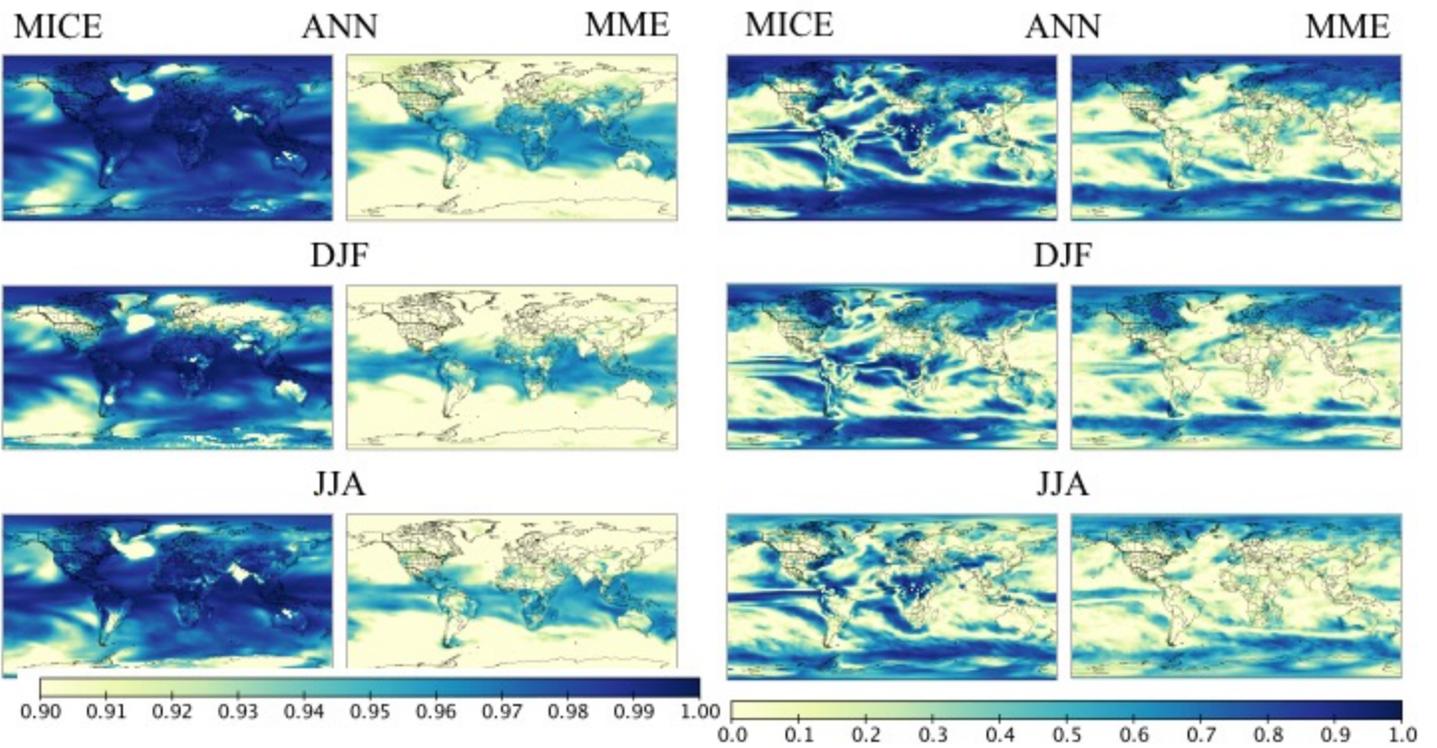

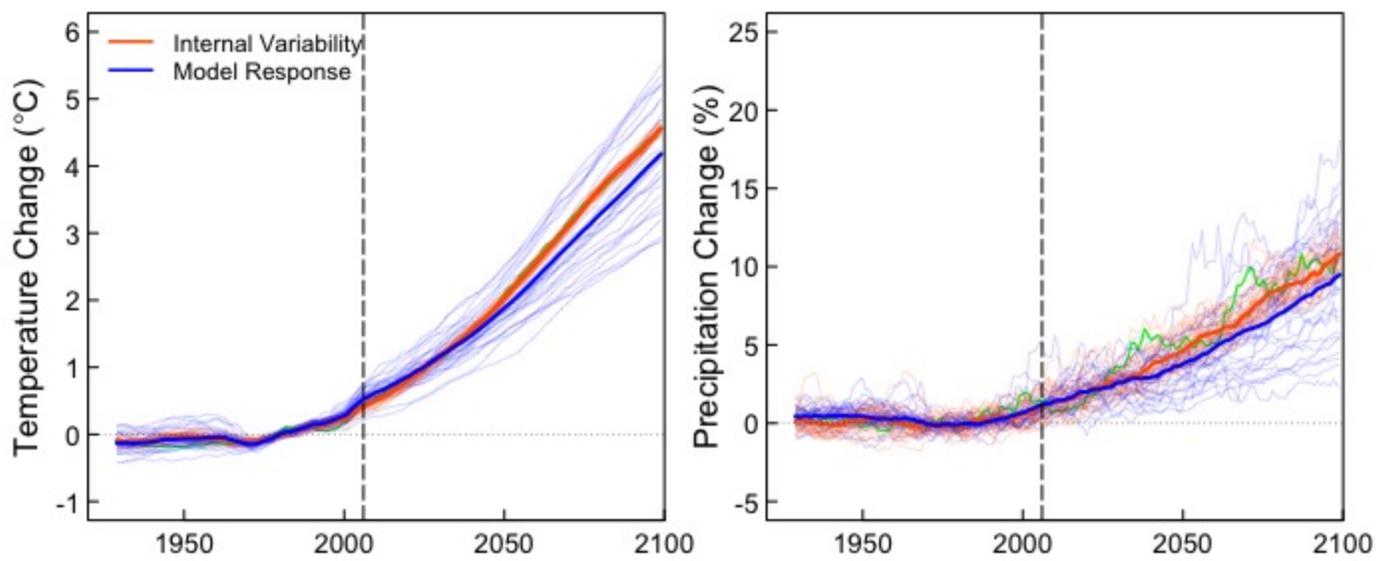
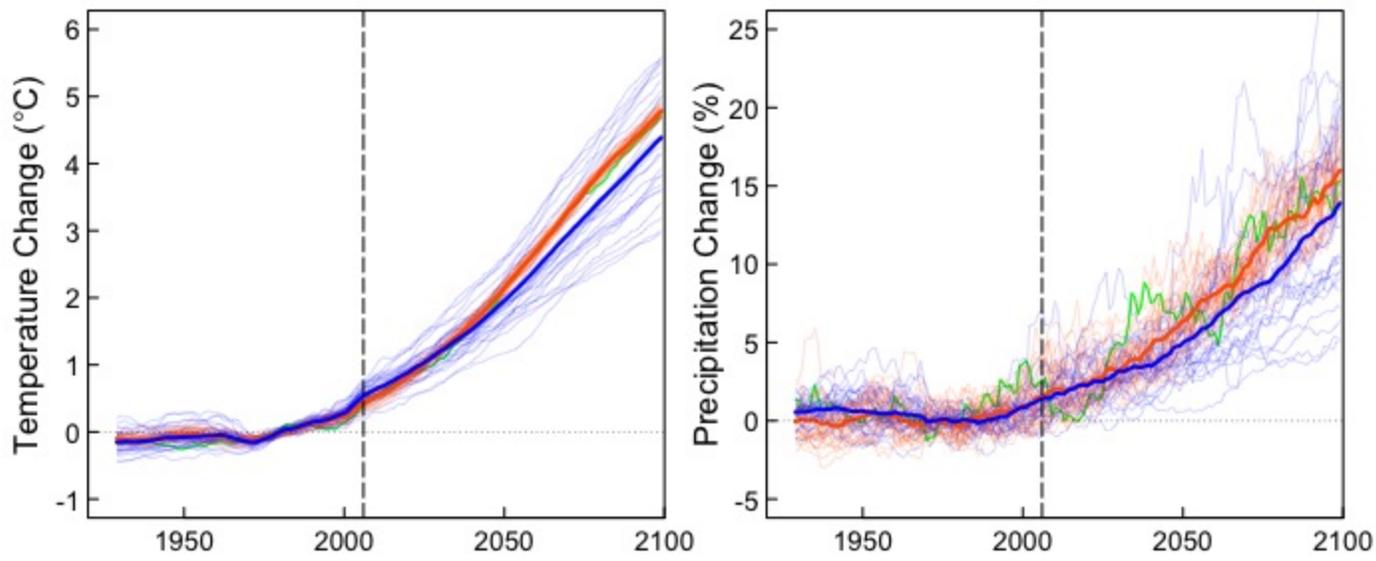
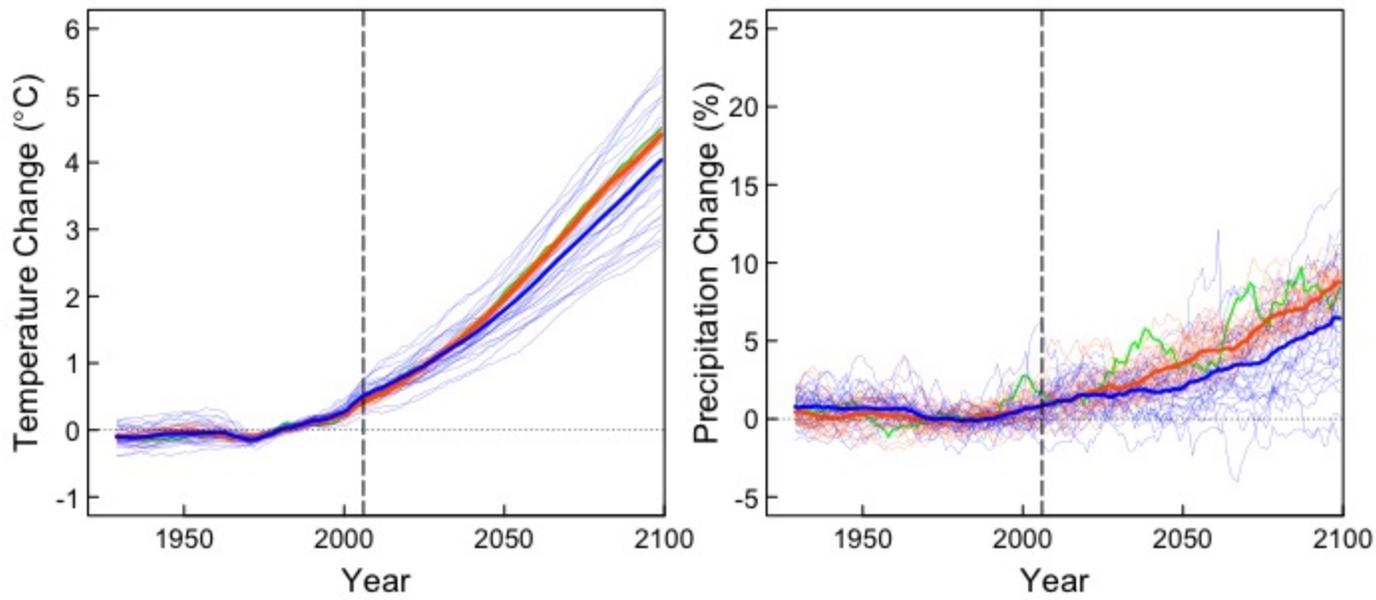

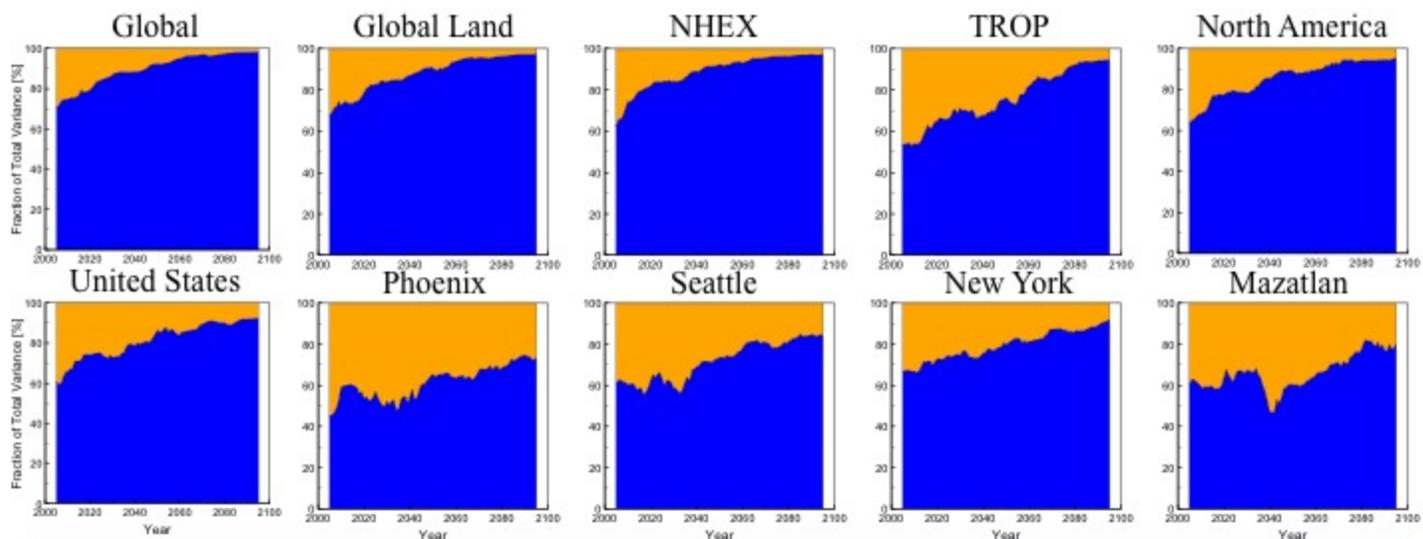
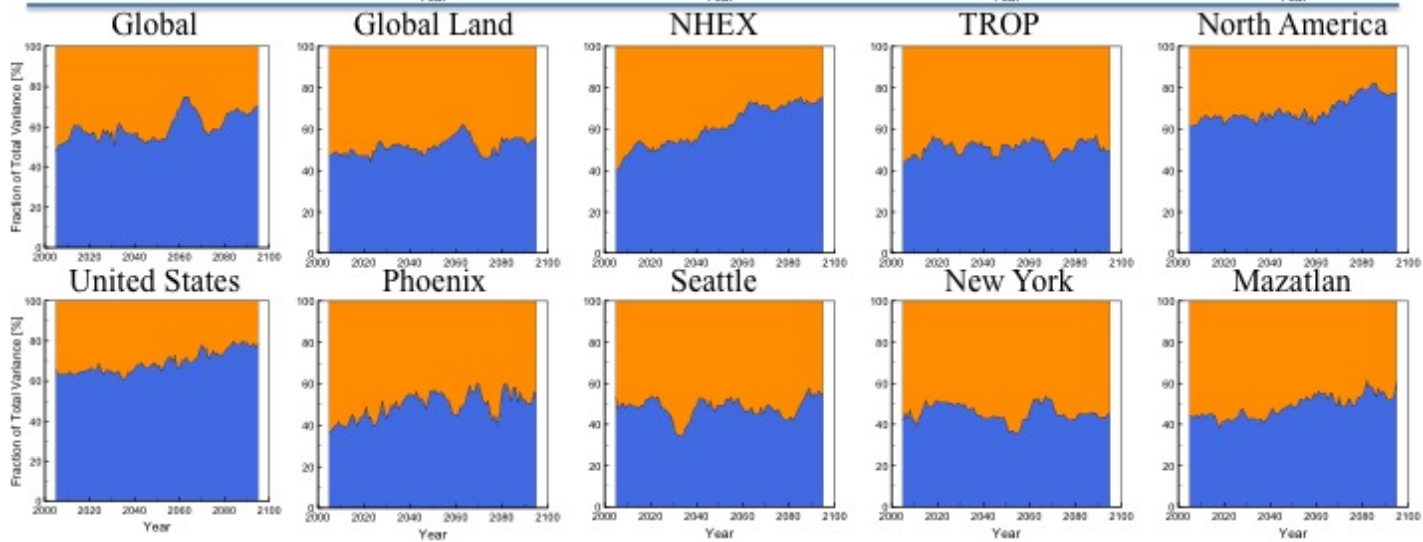

|            | Precipitation | Temperature |
|------------|---------------|-------------|
| 2010-2019  | 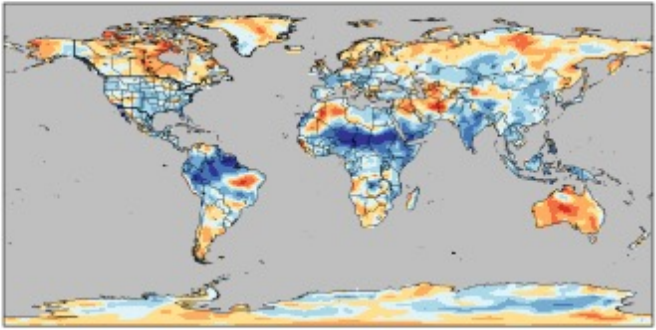 | 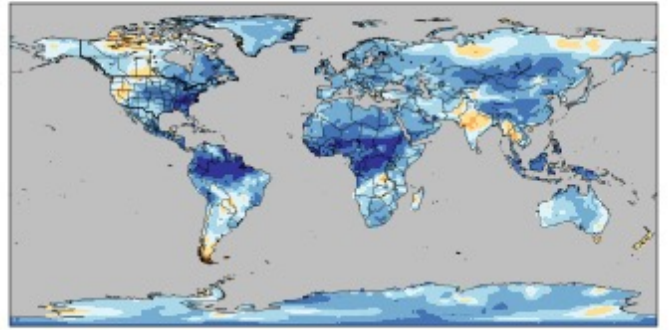 |
| 2030-2039  | 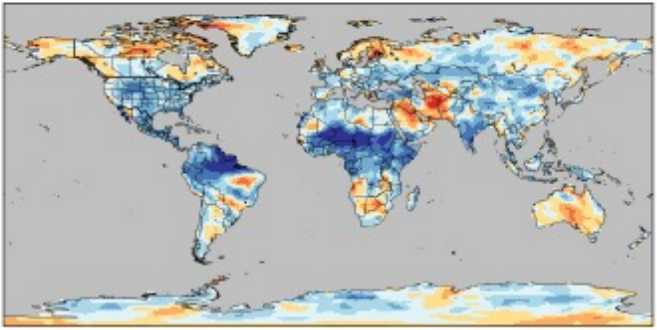 | 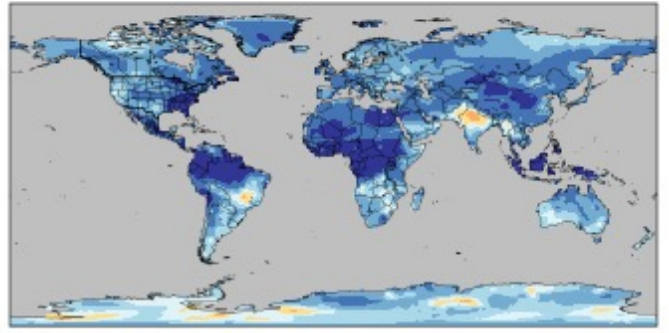 |
| 2050-2059  | 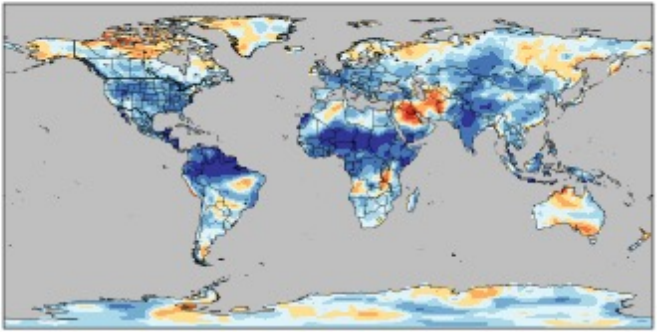 | 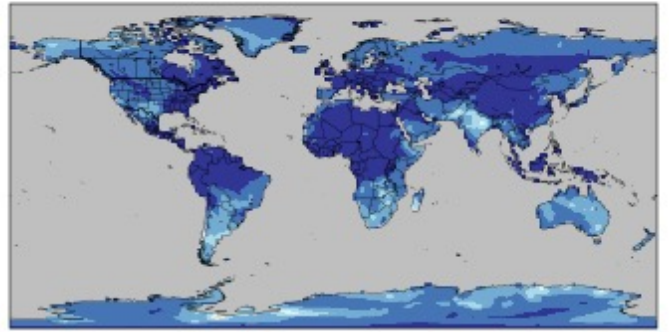 |
| 2090-2099  | 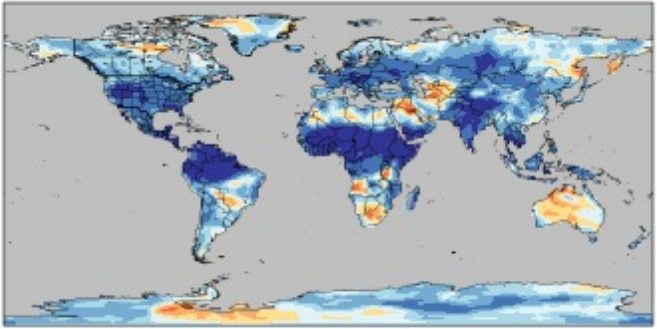 | 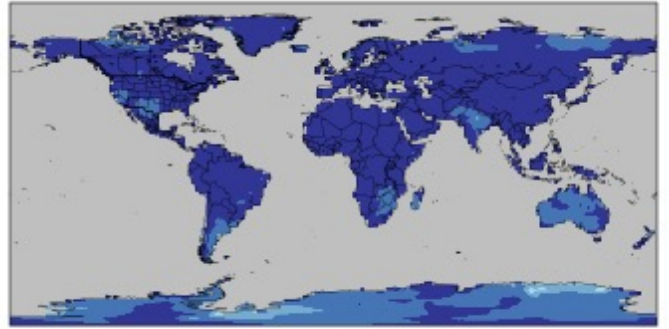 |

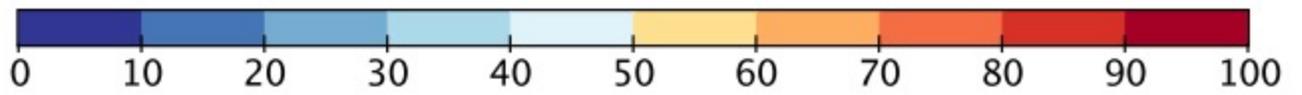